\documentclass[aps,prl,twocolumn,superscriptaddress,showpacs]{revtex4}
\usepackage{graphicx}
\usepackage{mathrsfs}
\usepackage{bm}
\usepackage{verbatim}
\begin{document}
\title{Stabilizing topological phases in graphene via random adsorption}
\author{Hua Jiang}
\email{jianghuaphy@gmail.com}
\affiliation{International Center for Quantum Materials, Peking University, Beijing 100871, China}
\author{Zhenhua Qiao}
\email{zhqiao@physics.utexas.edu}
\affiliation{Department of Physics, The University of Texas at
Austin, Austin, Texas 78712, USA}
\author{Haiwen Liu}
\affiliation{Institute of Physics, Chinese Academy of Sciences, Beijing 100190, China}
\author{Junren Shi}
\affiliation{International Center for Quantum Materials, Peking University, Beijing 100871, China}
\author{Qian Niu}
\affiliation{Department of Physics, The University of Texas at
Austin, Austin, Texas 78712, USA}\affiliation{International Center for Quantum Materials, Peking University, Beijing 100871, China}
\date{\today}
\begin{abstract}
We study the possibility of realizing topological phases in graphene with randomly distributed adsorbates. When graphene is subjected to periodically distributed adatoms, the enhanced spin-orbit couplings can result in various topological phases. However, at certain adatom coverages, the intervalley scattering renders the system a trivial insulator. By employing a finite-size scaling approach and Landauer-B\"{u}ttiker formula, we show that the randomization of adatom distribution greatly weakens the intervalley scattering, but plays a negligible role in spin-orbit couplings. Consequently, such a randomization turns graphene from a trivial insulator into a topological state.
\end{abstract}
\date{\today}
\maketitle
\textit{Introduction---} Due to its unusual structure, graphene displays two other internal degrees of freedom in addition to the real spin: AB sublattice and valley~\cite{AHCastro}. The former refers to the interpenetrating triangular sublattices A and B, while the latter refers to a pair of inequivalent Dirac cones $K$ and $K'$ in the Brillouin zone. When graphene is deposited on substrates or adsorbed with heavy atoms, the interaction-induced symmetry breaking can open bulk energy gaps to support a rich variety of topological phases. For example, a staggered AB sublattice potential breaking the inversion symmetry leads to a quantum valley-Hall effect~\cite{SiC-Gap,BN-Gap,ValleyHall}; certain nonmagnetic adatoms, i.e., indium and thallium atoms~\cite{Franz}, can enhance the intrinsic spin-orbit coupling (SOC) in graphene to host the quantum spin-Hall state~\cite{CLKane1}; some $3d$ or $5d$ transition metal adatoms produce a quantum anomalous-Hall state~\cite{haldane,HBZhang,ZHQiao1}, a consequence of the interplay between the proximity-induced magnetization and extrinsic Rashba SOC~\cite{ZHQiao2}. It is noteworthy that the intervalley scattering in these studies is completely avoided by choosing appropriate adatom coverages with valleys being separated in the momentum space, e.g., one adatom in each $4\times 4$ supercell of graphene (6.25\% adatom coverage).

However, at some specific adatom coverages, i.e., one adatom in each $3n\times3n$ ($n=1,2,3...$) supercell of graphene, valleys $K$ and $K'$ are folded into the $\Gamma$ point. The resulting intervalley scattering becomes significant and drives the system into a trivial insulator~\cite{ZHQiao2,3X3Gap}. In realistic graphene samples, the precise control of periodically distributed adatoms is impractical. The randomly distributed adatoms will inevitably cause mixture of different coverages and introduce the intervalley scattering. Therefore, a crucial issue arises: taking the site randomization of adatoms into account, is it possible to observe the topological phases experimentally?

In this Letter, we show that the randomization of adatom distribution greatly weakens the intervalley scattering, but does not affect the induced SOCs and magnetization, therefore increasing the possibility of realizing topological phases in graphene. By making use of a finite-size scaling method, we show that the randomization can induce a topological phase transition from a trivial insulator to a topologically nontrivial phase. Using the Landauer-B\"{u}ttiker formula, we confirm our finding by computing the two-terminal conductance in the presence of periodically or randomly distributed adatoms.

\textit{Quantum spin-Hall effect ---} We consider a graphene sheet adsorbed with diluted nonmagnetic atoms (e.g. indium or thallium), which prefer the hollow site of graphene~\cite{Franz}. For simplicity, we assume that the adatoms only interact with the surrounding six-nearest carbon atoms. Such an interaction not only enhances the intrinsic SOC, but also generates an on-site potential (also known as the crystal field stabilization energy) on each of the influenced carbon atoms. This potential is a key factor to induce the intervalley scattering.

In the presence of randomly distributed adatoms, the tight-binding Hamiltonian of graphene can be written as~\cite{CLKane1,Franz,Shevtsov}:
\begin{eqnarray}
H=&-& t \sum_{\langle i j \rangle,\alpha} c_{i\alpha}^{\dag}c_{j\alpha}+i \lambda_{\rm SO} \sum_{\langle\langle i j \rangle\rangle \in \mathcal{R} ,\alpha \beta} \nu_{ij}c_{i \alpha}^{\dag} s_{\alpha\beta}^{z} c_{j \beta} \nonumber \\
&+&U \sum_{i \in \mathcal{R},\alpha} c_{i\alpha}^{\dag} c_{i\alpha},\label{Equation1}
\end{eqnarray}
where $c_{i \alpha}^{\dag}$ creates an electron on site $i$ with spin $\alpha$, and $t$ is the hopping energy between nearest neighbors. The last two terms represent respectively the intrinsic SOC $\lambda_{\rm SO}$ and on-site energy $U$, which are applied on the influenced atomic sites denoted by $\mathcal{R}$. $s_z$ is the $z$-component of spin Pauli matrices. ${\langle \langle...\rangle \rangle}$ sums over all next-nearest neighbors. $\nu_{ij} =1(-1)$ corresponds to the hopping clockwise (counterclockwise) between next-nearest neighbors. According to their formation mechanisms, it is known that $\lambda_{\rm SO}$ should be one order smaller than $U$. In the following, we adopt the strength of intrinsic SOC in the thallium-atom adsorption case, i.e., $\lambda_{\rm SO}=0.016t \approx 0.044 $~eV ~\cite{Franz}, and the on-site energy is set to be $U=0.36t \approx 1.0$~eV.

As mentioned in the Introduction, the topological phases in graphene are sensitive to the coverage of periodically distributed adatoms. To make our investigation complete and convincing, we shall discuss the effect of randomization on two kinds of adatom coverage with and without intervalley scattering. In the following, using a finite-size scaling method, we simulate a fully random adsorption system by randomizing the distribution of adatoms in a gradually increasing supercell at the fixed adatom coverage. We begin with the discussion on the 11.1\% adatom coverage with intervalley scattering.
\begin{figure}
\includegraphics[width=8cm,  viewport=3 313 767 1092, clip]{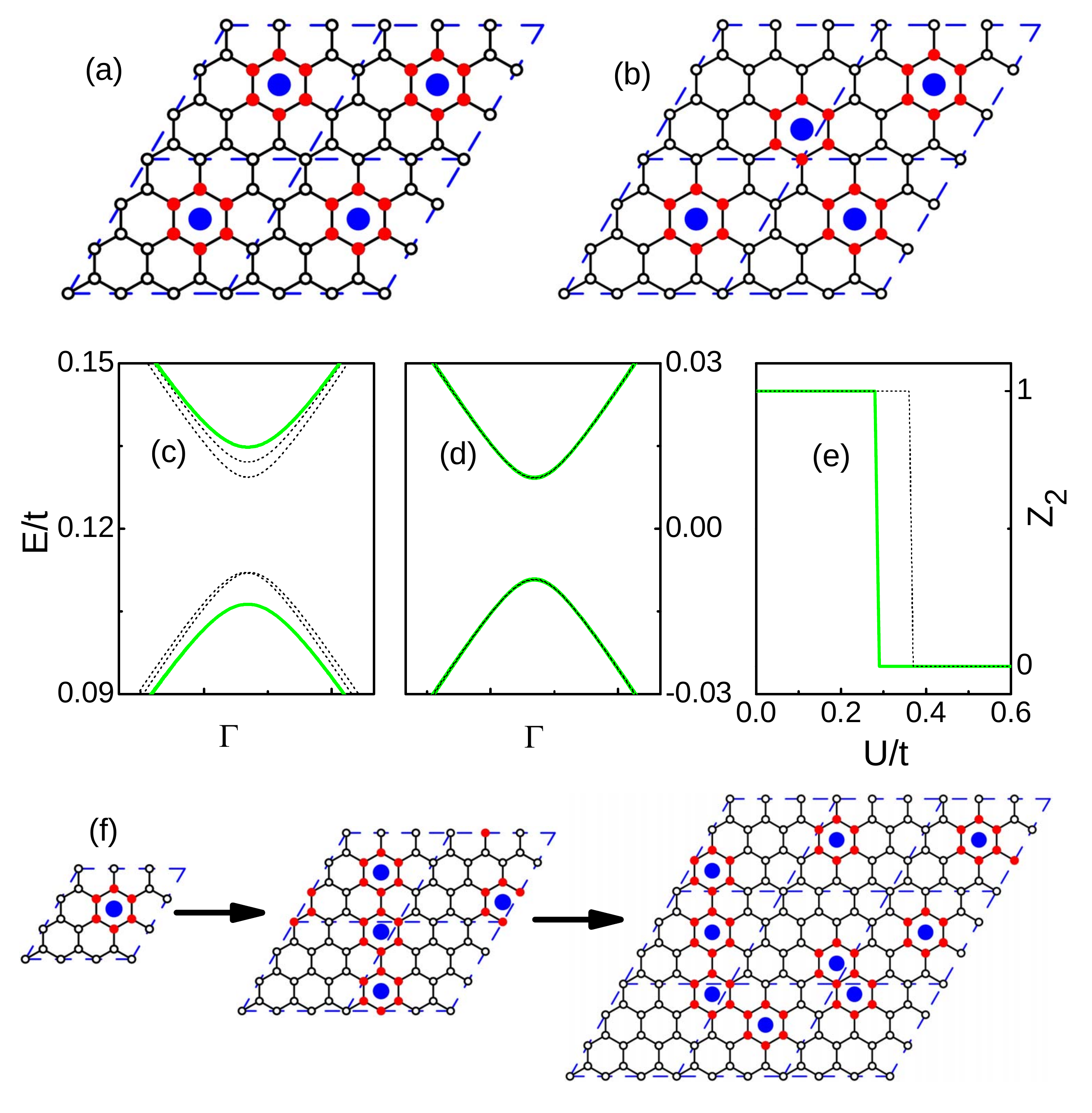}
\caption{(Color online) (a)-(b) Two different configurations of 4 adatoms in a $6 \times 6$ supercell. (c)-(d) Bulk band structure in the presence of only on-site energy $U=0.36t$ [panel (c)] or intrinsic SOC $\lambda_{\rm SO}=0.016t$ [panel (d)]. (e) $Z_2$ topological number versus $U$ at fixed $\lambda_{\rm SO}=0.016t$. In (c)-(e), solid and dashed lines correspond respectively to the supercells in (a) and (b). (f) Schematic of the finite-size scaling at a fixed $11.1\%$ adatom coverage. Empty, small solid and large solid circles represent the pristine graphene, carbons influenced by adatoms, and adatoms, respectively.}\label{Fig1}
\end{figure}

Let us first examine the sensitivity of topologically trivial/nontrivial phases to the adatom configuration. Figures \ref{Fig1}(a) and \ref{Fig1}(b) display two different $6\times6$ supercells of graphene. To keep the coverage, four adatoms are included in each supercell. In Fig.~\ref{Fig1}(a) the $3\times3$ periodicity still holds, while in Fig.~\ref{Fig1}(b) the $3\times3$ translational symmetry is broken due to the shift of one adatom to its neighboring site. This can be regarded as the simplest step to randomize a $3\times3$ supercell to a $6\times6$ one. To understand the role of randomization on the on-site potential $U$ and intrinsic SOC $\lambda_{\rm SO}$, we plot the bulk band structure around $\Gamma$ point by considering only $U$ or $\lambda_{\rm SO}$. In the presence of only $U$, the intervalley scattering in the $3\times3$ supercell opens a bulk gap as plotted in solid line in Fig.~\ref{Fig1}(c). One can see that a slight change of the adatom configuration shrinks the bulk band gap, implying the decrease of the invtervalley scatterng. As a sharp contrast, the band structures in the presence of only intrinsic SOC are almost identical for both supercells [see Fig.~\ref{Fig1}(d)]. This indicates that the effect of intrinsic SOC is insensitive to the slight change of the adatom configuration.

Although both on-site potential and intrinsic SOC open bulk gaps, the resulting insulators are topologically different. An efficient way to identify these phases is to calculate the $Z_2$ topological number, which characterizes the bulk band topology of a time-reversal invariant system~\cite{LiangFu}. By computing $Z_2$ topological number using the methods discussed in Ref.~\cite{Fukui}, we find that when both $U$ and $\lambda_{\rm SO}$ are nonzero, their competition results in different phases. As shown in Fig.~\ref{Fig1}(e), for a fixed $\lambda_{\rm SO}=0.016t$, small $U$ leads to a $Z_2=1$ quantum spin-Hall insulator, while large $U$ gives rise to a $Z_2=0$ trivial insulator. A remarkable difference between the two configurations can be observed in the range of $U \in [0.29t,~0.36t]$: a phase transition occurs from a trivial insulator to a $Z_2=1$ topological insulator.

\begin{figure*}
\includegraphics [width=16cm, viewport=38 178 1100 717, clip]{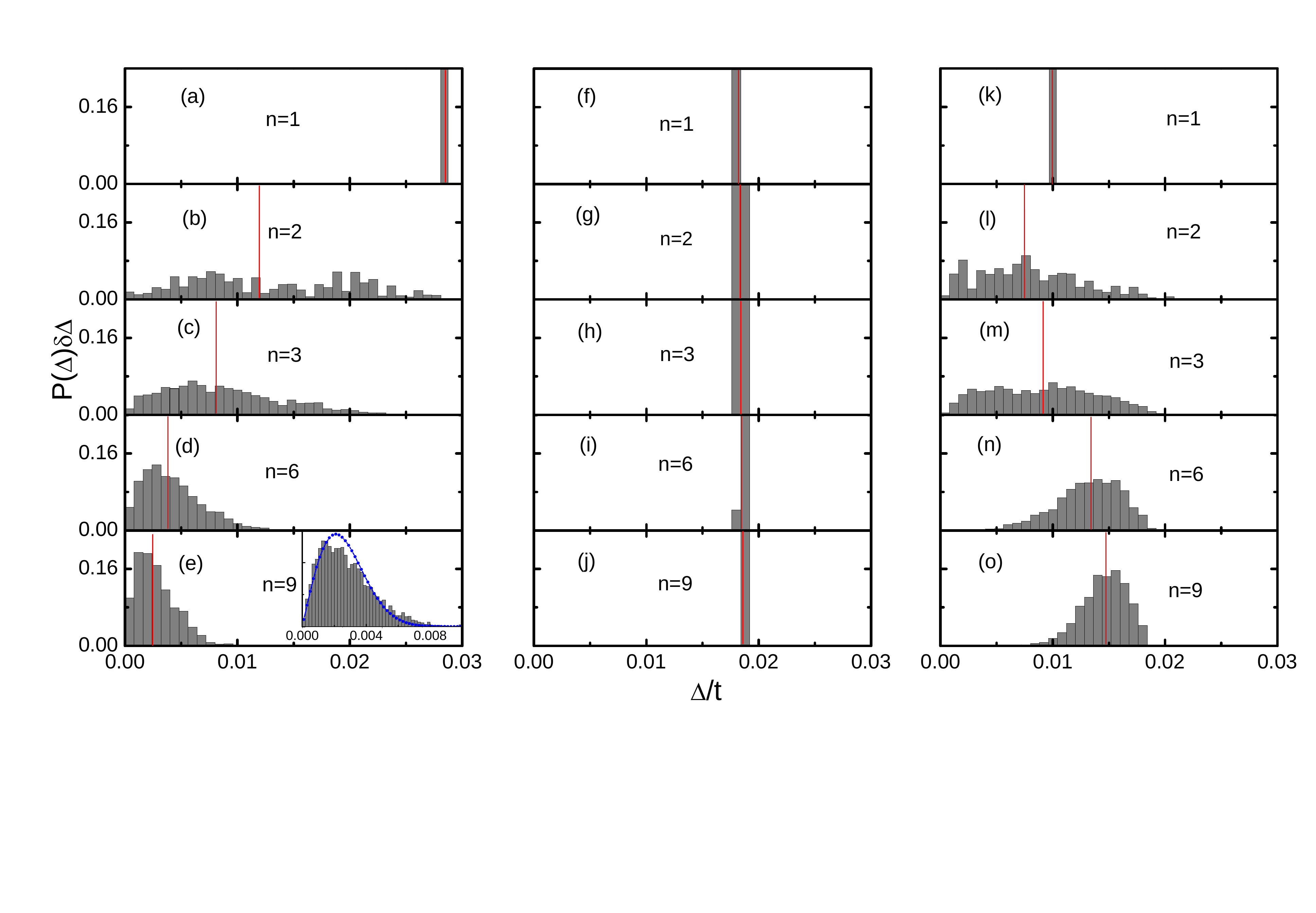}
\caption{(Color online) Probability distribution of bulk gap $P(\Delta)$ for $3n \times 3n$ supercells of graphene subjected to $n^2$ randomly distributed adatoms. $2~000$ samples are collected for each panel. (a)-(e) Only the crystal field stabilization energy $U=0.36t$ is considered. (f)-(j) Only the intrinsic SOC $\lambda_{\rm SO}=0.016t$ is considered. (k)-(o) Both $U=0.36t$ and $ \lambda_{\rm SO}=0.016t$ are included. The red line labels the median of each ensemble $\overline{\Delta}$. Inset: Numerical fitting of the probability distribution for $n=9$.}\label{Fig2}
\end{figure*}

Through employing a finite-size scaling method, we study the effect of strong randomization of adsorption sites. The numerical simulation procedures can be summarized as: (1) Expanding the supercell from $3 \times 3$ to $3n \times 3n$ ($n=2,3,4...$); (2) Randomly selecting $n^2$ hollow sites and then determining the influenced atomic sites $\mathcal{R}$; (3) Based on the Hamiltonian in Eq.~\ref{Equation1}, calculating the bulk band structure and measuring the bulk gap $\Delta$; (4) Repeating the procedures (2) and (3) to obtain $M$ samples. After that, the probability distribution of bulk gap $P(\Delta)$ can be obtained using the formula $P(\Delta) \delta \Delta = m/M$, where $m$ counts the magnitudes locating within the range of $[\Delta-\delta \Delta/2, \Delta+\delta \Delta/2)$.

Figure~\ref{Fig2} exhibits the evolution of the probability distribution of bulk gap $P(\Delta)$ along with the increasing of the supercell size $n$. The left column corresponds to the case with only the on-site potential $U=0.36t$. For $n=1$, although there are $C_9^1=9$ different adatom configurations, their band structures are exactly the same. Therefore, the resulting bulk gap is a constant $\Delta=0.029t$ [see Fig.~\ref{Fig2}(a)]. When $n>1$, there are $C_{9n^2}^{n^2}$ adatom configurations in a $3n\times3n$ supercell, e.g., a $9\times9$ supercell has $C_{81}^{9}=2.61 \times 10^{11}$ adatom configurations, most of which result in distinct band structures. In Figs.~\ref{Fig2}(b)-\ref{Fig2}(e), we observe that the band gap $\Delta$ fluctuates in a wide region, and the gap region shrinks towards zero for larger $n$. To better reflect this characteristic, we introduce the median of each ensemble $\overline{\Delta}$, which is highlighted in a red line. The exponential decay of $\overline{\Delta}$ as a function of $n$ can be easily visualized. In the inset of Fig.~\ref{Fig2}(e), we show that the probability distribution $P(\Delta)$ of $n=9$ can be fitted by a 2D Maxwell distribution function $f(\Delta)= {\Delta}/{\sigma^2} \exp (-{\Delta^2}/{2\sigma^2})$, where $\sigma= {\overline{\Delta}}/{\sqrt{2 \ln 2}}$. According to $f(\Delta)$, the probability of opening a band gap in the range of $[3 \overline{\Delta}, \infty)$ is about $0.2\%$.  Together with the fact that a realistic graphene sample resembles a $n\rightarrow \infty$ supcell, one can reasonably conclude that the intervalley scattering should be vanishing in the presence of randomly distributed adsorbates.

Next we turn to the case with only the intrinsic SOC $\lambda_{\rm SO}=0.016t$. As shown in the middle column, we find that the bulk gap $\Delta$ is almost independent of the randomization for any supercell size $n$. This means that the intrinsic SOC is still insensitive to the strong randomization of adatom distribution, and the resulting quantum spin-Hall phase is stable against the random adsorption. Comparing the results from both limits of only on-site potential or intrinsic SOC, it is natural to expect that the realistic graphene sample should favor the quantum spin-Hall state when both $U$ and $\lambda_{\rm SO}$ are present.

Such a speculation is verified by the gap statistics drawn in the right column. In the $3\times3$ supercell, because of the competition between $U$ and $\lambda_{\rm SO}$, the gap opening is smaller than those shown in Figs.~\ref{Fig2}(a) and \ref{Fig2}(f). When the supercell size increases, the median $\overline{\Delta}$ first decreases at $n=2$, but then increases for larger $n$ [see Figs.~\ref{Fig2}(m)-\ref{Fig2}(o)]. Through computing the $Z_2$ topological number, we obtain that $Z_2=0$ for $n=1$, while $Z_2=1$ for $n=3$, 6, and 9. Therefore, it is clear that the increasing of randomization turns graphene from a trivial insulator to a quantum spin-Hall insulator. From the tendency of $\overline{\Delta}$, one can notice that it gradually approaches the band gap labeled in Fig.~\ref{Fig2}(f).

\begin{figure*}
\includegraphics [width=16.5cm, viewport=11 220 1110 755, clip]{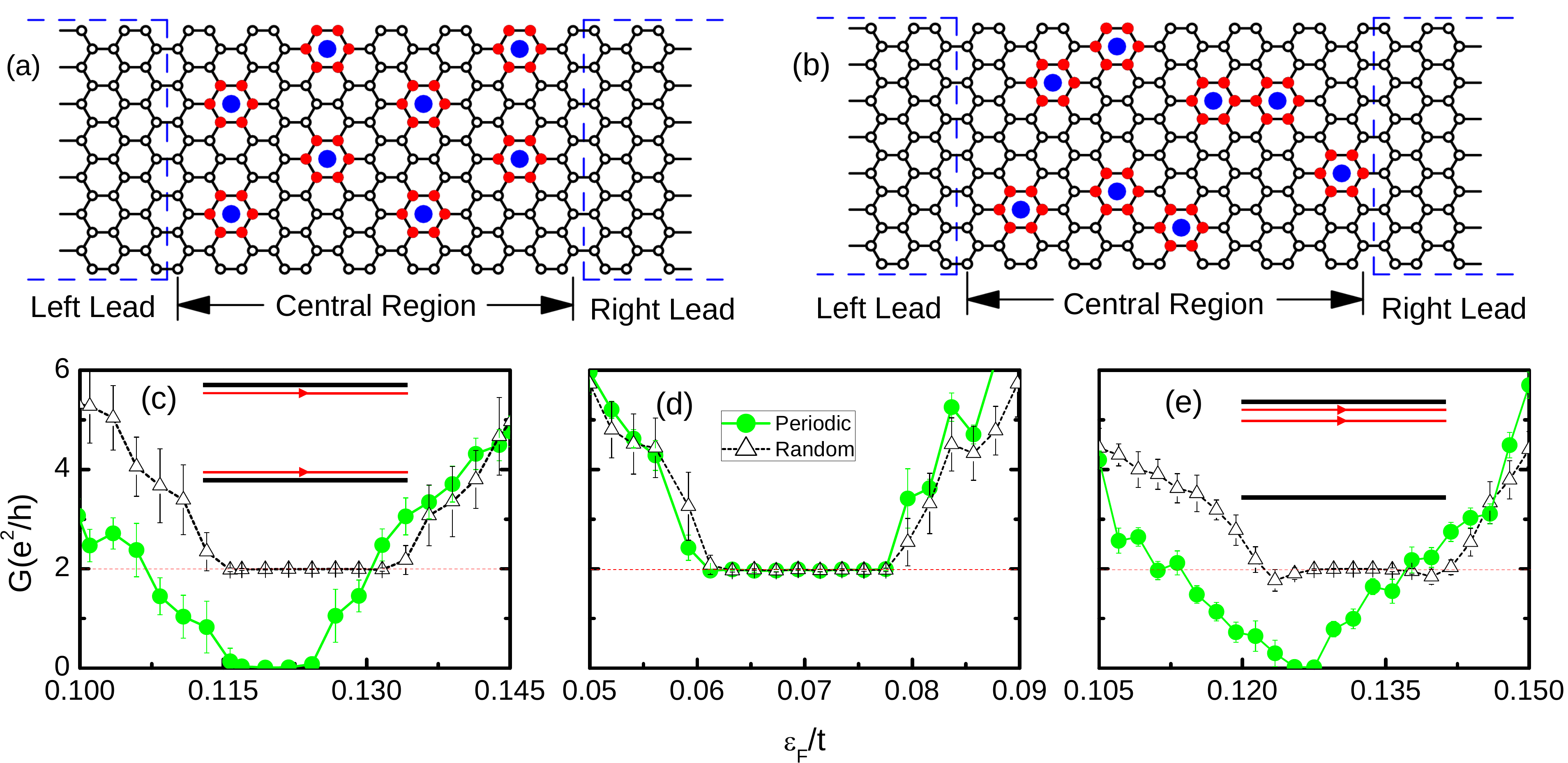}
\caption{(Color online) (a)-(b) Schematic of a two-terminal setup in the presence of periodically and randomly distributed adatoms, respectively. And both adatom coverages are 11.1\%. (c)-(e) Comparison of conductances $G$ between periodic and random adsorption as a function of Fermi level $\varepsilon_F$. (c)-(d) The on-site potential and intrinsic SOC are set to be $U=0.36t$ and $\lambda_{\rm SO}=0.016t$. And the adatom coverages are 11.1\% in panel (c) and 6.25\% in panel (d). In (e), the on-site potential, Rashba SOC, and exchange field are set to be $U=0.38t$, $\lambda_{R}=0.04t$ and $M=0.04t$, respectively. And the adatom coverage is 11.1\%. Circle and triangle symbols represent the periodic and random adatoms. The associated error bar denote the conductance fluctuation $\delta G$. Insets of (c) and (e): schematic of the quantized channels from left to right.}\label{Fig3}
\end{figure*}

Thus far, we analyze the influence of randomization by investigating the bulk band structures of various supercells. Hereinbelow, we design a two-terminal device illustrated in Fig.~\ref{Fig3} to study the transport properties using the Landauer-B\"{u}ttiker formula~\cite{SDatta}. The adatoms are only considered in the central scattering regime, and the two leads are modeled by pristine graphene ribbons. The presence of helical edge modes is one of the most striking properties of the quantum spin-Hall effect, which gives rise to a quantized longitudinal conductance.

Figure~\ref{Fig3}(c) plots the average conductance $G$ and its fluctuation $\delta G$ as a function of the Fermi energy $\varepsilon_F$. The parameters are $U=0.36t$ and $\lambda_{SO}=0.016t$. In the presence of periodically distributed adatoms [see Fig.~\ref{Fig3}(a)], $G=0$ and $\delta G =0$ in units of $e^2/h$ within the range of $\varepsilon_F\in [0.117t,~0.124t]$, signaling a trivial insulator. However, when the adatoms become randomly distributed in Fig.~\ref{Fig3}(b), a quantized plateau $G=2e^2/h$ with vanishing fluctuation emerges in the regime of $\varepsilon_F \in[0.116t,~ 0.132t]$. The distribution of local currents illustrated in the inset of (c) further indicates a quantum spin-Hall insulator. Such a phase transition resembles the disorder-induced topological Anderson insulator~\cite{JianLi,Carlo}. The major difference is that it is the adatom configuration but not the disorder strength to trigger the phase transition.

Throughout the above analysis, we focus only on the 11.1\% adatom coverage with strong intervalley scattering. What happens for other coverages without intervalley scattering? Let us take the 6.25\% adatom coverage (one adatom in a $4\times4$ supercell) for example. Using the same finite-size scaling method, we show that the bulk gap is only dependent on the intrinsic SOC, but independent of the on-site energy or supercell size. An immediate evidence is the robust quantized plateau shown in Fig.~\ref{Fig3}(d) for either periodically or randomly distributed adatoms. To conclude, in a thallium-atom adsorbed graphene, the quantum spin-Hall state is a system-preferred ground state for any adatom coverage, and the bulk gap $\sim46$~meV can be detected under current experimental techniques.

\textit{Quantum anomalous Hall effect---} When the $3d/5d$ transition metal atoms are adsorbed on graphene, the interaction enhances the Rashba SOC $\lambda_R$~\cite{note}, and induces the on-site potential $U$ and magnetization $M$. In Refs.~\cite{HBZhang} and \cite{ZHQiao2}, it is found by \emph{ab initio} calculations  that quantum anomalous Hall phase can be produced in the 6.25\% adatom coverage, but the trivial insulator is usually formed in the 11.1\% adatom coverage due to the strong intervalley scattering. In Fig.~\ref{Fig3}(e), we calculate the average conductance versus the Fermi energy $\varepsilon_F$ at a 11.1\% adatom coverage. The parameters are chosen to be $U=0.38t$, $\lambda_R=0.04t$, and $M=0.04t$. We observe that $G=0e^2/h$ in the range of $\varepsilon_F \in [0.125t,~0.127t]$ for the periodically distributed adatoms. However, for the randomly distributed adatoms, a quantized plateau $G=2e^2/h$ emerges in the range of $ \varepsilon_F \in [0.127t,~0.137t] $. The inset plots the schematic of the chiral edge modes from left to right, which indicates a quantum anomalous Hall insulator. This result further confirms our finding that the intervalley scattering is fragile, but the spin-orbit couplings and exchange field are robust against the randomization of adatom distribution.

\textit{Summary---} In the presence of periodically distributed adatoms, the interaction induced spin-orbit couplings can produce various topological phases when the intervalley scattering is completely suppressed. At certain adatom coverage, the strong intervalley scattering plays a dominant role and turns graphene into a trivial insulator. Using both finite-size scaling method and transport calculation, we show that when the adatom distribution becomes random, the intervalley scattering is weakened, but other quantities (e.g. spin-orbit couplings, and exchange field) are not affected. This finding points out that the topological states are graphene-favored ground states in the presence of randomly distributed adtoms.

\textit{Acknowledgements---} We are grateful to Y. G. Yao, J. H. Zhou, Q. F. Sun, and X. C. Xie for valuable discussions. H.J. was supported by China Postdoctroal Science Foundation (20100480147 and 201104030) and the MOST Project of China (2012CB921300). Z.Q. was supported by NSF (DMR 0906025) and Welch Foundation (F-1255). Q.N. was supported by DOE (DE-FG03-02ER45958, Division of Materials Science and Engineering).

\end{document}